\newcommand{\be}{\begin{equation}}
\newcommand{\ee}{\end{equation}}
\newcommand{\nl}{\nonumber \\}

\newcommand{\psibar}{\overline{\Psi}}

\newcommand{\ewxy}[2]{\setlength{\epsfxsize}{#2}\epsfbox[0  0  880  700 ]{#1}}

\documentclass{PoS}
\usepackage{epsf}

\title{Recent results on B mixing and decay constants from HPQCD}

\ShortTitle{Recent results on B mixing and decay constants from HPQCD}

\author{\speaker{J.~Shigemitsu}$^a$, C.~Davies$^b$, E.~Follana$^c$,
        E.~G\'amiz$^d$, E.~Gregory$^b$, P.~Lepage$^e$, H.~Na$^a$ 
and M.~Wingate$^f$\\
\llap{$^a$}Department of Physics \\
             The Ohio State University, Columbus, OH 43210, USA\\
\llap{$^b$}Department of Physics \& Astronomy \\
           University of Glasgow, Glasgow G12 8QQ, UK\\
\llap{$^c$}Departamento de F\'isica Te\'orica\\
           Universidad de Zaragoza, E-50009 Zaragoza, Spain\\
\llap{$^d$}Department of Physics\\
           University of Illinois, Urbana, IL 61801, USA\\
\llap{$^e$}Laboratory of Elementary Particle Physics\\
           Cornell University, Ithaca, NY 14853, USA\\
\llap{$^f$}Department of Applied Mathematics and Theoretical Physics\\
           University of Cambridge, Cambridge CB3 0WA, UK\\
        E-mail: \email{shige@mps.ohio-state.edu}}

\abstract{We review recent results for $B_d$ and $B_s$ mixing parameters 
using MILC $N_f = 2 + 1$ lattices, NRQCD b-quarks and AsqTad light quarks.  
Latest numbers for decay constants $f_B$ and $f_{B_s}$ are also 
presented. Combining our lattice results with experimental determinations 
of the mass differences $\Delta M_d$ and $\Delta M_s$ leads to an 
important ratio of elements of the CKM matrix,
$|V_{td}|/|V_{ts}| = 0.214(1)(5)$ and an updated Standard Model number 
for the branching fraction $Br(B_s \rightarrow \mu^+\mu^-) = 
3.19(19) \times 10^{-9}$. Preliminary new results for 
$f_{B_s}$ based on other actions are also described. }

\FullConference{The XXVII International Symposium on Lattice Field Theory\\
                 July 26-31, 2009\\
                 Peking University, Beijing, China}

\begin{document}

\section{Introduction}
$B$ Physics, and in particular the study of $B$ and $B_s$ meson decays 
and mixing, remains an important part of Flavor Physics.  Such studies 
enable consistency tests of the Standard Model and  placing of 
bounds on New Physics effects. Lattice QCD is playing a crucial role 
in this effort by providing the necessary nonperturbative QCD inputs.

 This talk describes recent results by the HPQCD collaboration 
on B meson mixing parameters, $f_{B_s} \sqrt{\hat{B}_{B_s}}$, 
$f_{B_d} \sqrt{\hat{B}_{B_d}}$ and their ratio $\xi$, 
 the first fully consistent 
$N_f = 2 +1$ lattice QCD calculations of these quantities \cite{bmix}.
We work with 
four of the MILC coarse ($a \approx 0.12$fm) lattices and two of the 
fine ($a \approx 0.09$fm) lattices. We use NRQCD for the $b$-quark and 
the AsqTad action  for both valence and sea light quarks.  
Details are given in \cite{bmix}.  In the Standard Model 
the mass difference in the $B_q - \overline{B_q} \;$ 
($q = d, s$) system is given by \cite{buras},
\begin{equation}
\label{deltamq}
\Delta M_q = \frac {G_F^2 M_W^2}{6 \pi^2} |V_{tq}V^*_{tb}|^2 \eta_2^B
S_0(x_t) M_{B_q} f^2_{B_q} \hat{B}_{B_q}.
\end{equation}
In addition to well determined quantities, eq.(\ref{deltamq}) involves 
the combinations of CKM matrix elements $|V_{tq} V^*_{tb}|^2$ and 
the nonperturbative QCD factors $f^2_{B_q} \hat{B}_{B_q}$.
The latter are determined from the matrix element of the 
four-quark operator 
 ($i,j$ are color indices that are summed over),
\begin{equation}
\label{ol}
OL \equiv \left [ \overline{\Psi}^i_b (V - A)\Psi^i_q \right ]
\; \left [ \overline{\Psi}^j_b (V - A)\Psi^j_q \right ]
\end{equation}
sandwiched between the $B_q$ and $\overline{B}_q$ states.
\be
\label{defol}
\langle OL \rangle ^{\overline{MS}}(\mu)
\equiv
\langle \overline{B}_q | OL| B_q \rangle ^{\overline{MS}}(\mu)
\equiv \frac{8}{3}
f^2_{B_q} \, B_{B_q}(\mu)\,  M^2_{B_q}.
\ee
One sees from eq.(\ref{deltamq}) that the ratio of the two CKM matrix 
elements $|V_{td}|$ and $|V_{ts}|$ can be determined from 
\be
\frac{|V_{td}|}{|V_{ts}|} = \xi \sqrt{\frac{\Delta M_d}{\Delta M_s}
 \frac{M_{B_s}}{M_{B_d}}} ,
\ee
once  theory provides the important ratio
\be
\xi \equiv \frac{f_{B_s}
\sqrt{B_{B_s}}}{f_{B_d} \sqrt{B_{B_d}}}.
\ee
One of the main goals of \cite{bmix} was to produce a
state-of-the-art lattice calculation of $\xi$.  We were able to determine 
this quantity with total error of 2.6\%.

\section{Hadronic Matrix Elements in the Effective Theory}
The $b$ quarks in our calculations are described by an 
effective theory NRQCD.  Instead of the full QCD field $\psibar_b$ 
of eq.(\ref{ol}) 
one works with fields $\psibar_Q$ that create a heavy quark or
 $\psibar_{\overline{Q}}$ that
 annihilate a heavy anti-quark.  The $[V-A] \times [V-A]$ 
four-fermion operator becomes,
\be
OL^{eff}   \equiv  \left[\psibar_Q^i(V-A) \Psi_q^i \right]\,
\left [\psibar_{\overline{Q}}^j (V-A) \Psi_q^j \right]  
+  \left[\psibar_{\overline{Q}}^i(V-A) \Psi_q^i \right]\,
\left [\psibar_Q^j (V-A) \Psi_q^j \right].
\ee
In the effective theory one finds mixing with another operator at 
one-loop of $[S-P] \times [S-P]$ structure,
 even at lowest order in $1/M$.
\be
OS^{eff}  \equiv  \left[\psibar_Q^i(S-P) \Psi_q^i \right]\,
\left [\psibar_{\overline{Q}}^j (S-P) \Psi_q^j \right] 
+  \left[\psibar_{\overline{Q}}^i(S-P) \Psi_q^i \right]\,
\left [\psibar_Q^j (S-P) \Psi_q^j \right].
\ee
Furthermore there is a tree-level dimension 7 correction to 
$OL^{eff}$ at ${\cal O}(1/M)$.
\begin{eqnarray}
\label{effj1}
OLj1 &=& \frac{1}{2M} \left [ \left (
 \vec{\nabla} \psibar_Q  \, \cdot \, \vec{\gamma} \, (V-A) \, \Psi_q \right )
 \left ( \psibar_{\overline{Q}} \, (V-A) \, \Psi_q \right ) \right . \nl
 &&  \; +  \quad \left . \left (
 \psibar_Q \, (V-A) \, \Psi_q \right )
 \left ( \vec{\nabla}\psibar_{\overline{Q}}
\, \cdot \, \vec{\gamma} \, (V-A) \, \Psi_q \right ) \right ] 
\quad + \quad 
\left [\psibar_{\overline{Q}} \rightleftharpoons \psibar_Q \right ] .
\end{eqnarray}
We have calculated the matching between matrix elements of the 
effective theory operators $OL^{eff}$, $OS^{eff}$ and $OLj1$ and 
$\langle OL \rangle$ of eq.(\ref{ol}) in full QCD 
through ${\cal O}(\alpha_s, \Lambda_{QCD}/M, \alpha_s/(aM))$ and 
find
\begin{eqnarray}
\label{olmsbar}
\langle OL \rangle^{\overline{MS}}(\mu) &=& 
 [ \, 1 + \alpha_s \, \rho_{11} \,]
 \, \langle OL^{eff} \rangle
  \, + \, \alpha_s \, \rho_{12} \, \langle OS^{eff} \rangle +  \nl
&  & \langle OLj1 \rangle
- \alpha_s \, \left [\, \zeta^{11} \,
\langle OL^{eff} \rangle \, + \,  \zeta^{12} \,
\langle OS^{eff} \rangle \, \right ] 
 \; \; + \;{\cal O}(\alpha_s^2, \alpha_s \Lambda_{QCD} / M).
\end{eqnarray}
The matching coefficients $\rho_{11}$, $\rho_{12}$, $\zeta^{11}$ and
$\zeta^{12}$ are listed (for $\mu = M_b$) in \cite{fourfmatch}.

The goal of lattice simulations is to obtain the matrix elements 
$\langle \hat{O} \rangle$, with $\hat{O} = OL^{eff}$, $OS^{eff}$ or
$OLj1$. To this end one calculates three-point correlators,
\be
\label{thrpnt}
 C^{(4f)}_{\alpha \beta}(t_1,t_2) = \nl
 \sum_{\vec{x}_1,\vec{x}_2} \langle 0 |
\Phi^{\alpha}_{\overline{B}_q}(\vec{x}_1,t_1) \; [a^6 \hat{O}(0)]
\; \Phi^{\beta \dagger}_{B_q}(\vec{x}_2,-t_2) | 0 \rangle, 
\ee
together with two-point correlators
\be
\label{twopnt}
 C^{2pt}_{\alpha \beta}(t)  \equiv   \sum_{\vec{x}_1,\vec{x}_2} \langle 0 |
\Phi^{\alpha}_{B_q}(\vec{x}_1,t)
\; \Phi^{\beta \dagger}_{B_q}(\vec{x}_2,0) | 0 \rangle .
\ee
One works with dimensionless operators $ a^6 \hat{O}$ which are
kept fixed at the origin of the lattice.
 $\Phi^{\alpha}_{B_q}$ is an interpolating operator for the $B_q$ meson
 of smearing type ``$\alpha$'', and
spatial sums over $\vec{x}_1$ and $\vec{x}_2$ ensure one is dealing with
zero momentum $B_q$ and $\overline{B_q}$ incoming and outgoing states.
Eq.(\ref{thrpnt}) corresponds to a $B_q$ meson (or excitations thereof with 
the same quantum numbers) being created at time $-t_2$ which then 
propagates to time $0$ where it mixes onto its anti particle a 
$\overline{B_q}$.  The latter meson is annihilated at time $t_1$.

We have accumulated simulation data for a range in $t_1,t_2$ for three-point
and in $t$ for two-point correlators, i.e. $1 \leq t_1,t_2,t \leq T_{max}$ 
with $T_{max} = 24$ on the coarse lattices and $T_{max} = 32$ on the 
fine ensembles. As is well known, staggered quarks lead to meson 
two-point correlators that have both regular and time oscillating 
contributions. Hence a fit ansatz for eq.(\ref{twopnt}) would be 
\be
\label{twopntfit}
 C^{2pt}_{\alpha \beta}(t)  \equiv 
\sum_{j=0}^{N-1} b^\alpha_j b^\beta_j e^{-E_j (t-1)}
 + (-1)^t \sum_{k=0}^{\tilde{N}-1}
\tilde{b}^\alpha_k \tilde{b}^\beta_k e^{-\tilde{E}_k (t-1)}. 
\ee
Similarly, the presence of oscillatory contributions makes fitting 
the three-point correlators, eq.(\ref{thrpnt}) particularly 
challenging. The appropriate ansatz is
\begin{eqnarray}
\label{cf4f}
 && C^{(4f)}_{\alpha \beta}(t_1,t_2) =
 \;\;\; \sum_{j=0}^{N-1} \sum_{k=0}^{N-1}  A^{\alpha \beta}_{jk} \;
 e^{-E_j (t_1-1)}\; e^{-E_k (t_2-1)} 
+ \sum_{j=0}^{\tilde{N}-1} \sum_{k=0}^{N-1}  B^{\alpha \beta}_{jk} \;
 (-1)^{t_1} \; e^{-\tilde{E}_j (t_1-1)}\; e^{-E_k (t_2-1)} \nl
&&  \qquad 
+ \sum_{j=0}^{N-1} \sum_{k=0}^{\tilde{N}-1}  C^{\alpha \beta}_{jk} \;
 (-1)^{t_2} \; e^{-E_j (t_1-1)}\; e^{-\tilde{E}_k (t_2-1)} 
+ \sum_{j=0}^{\tilde{N}-1} \sum_{k=0}^{\tilde{N}-1}  D^{\alpha \beta}_{jk} \;
 (-1)^{t_1} (-1)^{t_2}\; e^{-\tilde{E}_j (t_1-1)}\;
 e^{-\tilde{E}_k (t_2-1)}.  \nl
\end{eqnarray}
In terms of the fit parameters in (\ref{twopntfit}) and (\ref{cf4f}) 
the matrix elements we are interested in and which appear on the RHS of  
(\ref{olmsbar}) are given by,
\be
 \langle \overline{B_q} | \hat{O}| B_q \rangle = \frac{2 M_{B_q}}{a^3}
\frac{A^{\alpha \beta}_{00}}{b^\alpha_0 b^\beta_0}.
\ee
 In order to extract 
$\frac{A_{00}^{\alpha \beta}}{b^\alpha_0 b^\beta_0}$, 
we have carried out simultaneous fits to $C^{(4f)}_{\alpha \beta}
(t_1,t_2)$ for $\alpha = \beta$ together with a matrix of 
two-point correlators $C^{2pt}_{\alpha \beta}(t)$ with all possible 
combinations of smearings at source and sink.
We used Bayesian fitting methods and typically employed 
$N = \tilde{N} = 4 \sim 6$ number of exponentials.

\section{Chiral and Continuum Extrapolations}
After the fits to simulation data described above 
 (see \cite{bmix} for more details) one ends up with lattice 
results for $r_1^{3/2} f_{B_q} \sqrt{M_{B_q} \hat{B}_{B_q}}$ for 
each of the 6 MILC ensembles that we worked with. Here $r_1$ is a scale 
derived from the static potential which can be used to make dimensionful 
quantities dimensionless. We exploit the fact that the MILC collaboration 
has calculated $r_1/a$ for each of their ensembles.  In order to make 
contact with the real world one needs to extrapolate both to the 
continuum (lattice spacing $a \rightarrow 0$) and to the 
chiral ($m_{light} \rightarrow m_{u/d}$) limits. We do so by fitting 
to the following ansatz, inspired by chiral perturbation theory (ChPT).
\be
\label{schpt1}
 r_1^{3/2} \, f_{B_q}\sqrt{M_{B_q} \hat{B}_{B_q}} =  
   c_1 \, [1 +  \frac{1}{2} \,\Delta f_q   +
c_2 \, (2 m_f + m_s) \, r_1 +  c_3 \, m_q \, r_1 ] \times  
 [ 1 + c_4 \,
 \alpha_s ( a/r_1 )^2  + c_5 \,  ( a/r_1 )^4 ] .
\ee
$\Delta f_q$ includes the chiral logarithms including those specific 
to Staggered ChPT and was calculated by C.Bernard, J.Laiho
and R.Van de Water \cite{blv}.
We show  chiral/continuum extrapolations  for 
$\;r_1^{3/2} f_{B_s}\sqrt{M_{B_s}\hat{B}_{B_s}}$ and 
$\;\;r_1^{3/2} f_{B_d}\sqrt{M_{B_d}\hat{B}_{B_d}}$ in Fig.1 and for
$\;\;\xi \sqrt{M_{B_s}/M_{B_d}}$ and for $\;\hat{B}_{B_s}$ 
in Fig.2. 
The red 
curves are the continuum extrapolated curves and the red circle gives 
the results at the physical point.
Table I gives our error budget.  

\begin{table}[b]
\begin{center}
\begin{tabular}{|c|c|c|c|}
\hline
source of error & $f_{B_s} \sqrt{\hat{B}_{B_s}}$ &
 $f_{B_d} \sqrt{\hat{B}_{B_d}}$ & $\;\;\xi \;\;$ \\
\hline
\hline
stat. + chiral extrap.  & 2.3  &  4.1 &  2.0 \\
residual  $a^2$ extrap. &3.0  & 2.0 &  0.3\\
uncertainty &&&  \\
\hline
 $r_1^{3/2}$ uncertainty  & 2.3  &  2.3  &  --- \\
 $g_{B^*B\pi}$ uncertainty  & 1.0  &  1.0  &  1.0 \\
$m_s$ and $m_b$ tuning &  1.5  & 1.0  & 1.0  \\
operator matching  & 4.0 & 4.0& 0.7 \\
relativistic corr.  & 2.5 &  2.5 & 0.4\\
\hline
Total  & 6.7 & 7.1 & 2.6 \\
\hline
\end{tabular}
\end{center}
\caption{ Errors in \% for
$f_{B_s}\sqrt{\hat{B}_{B_s}}$, $f_{B_d}\sqrt{\hat{B}_{B_d}}$ and $\xi$.
}
\end{table}


\begin{figure}[t]
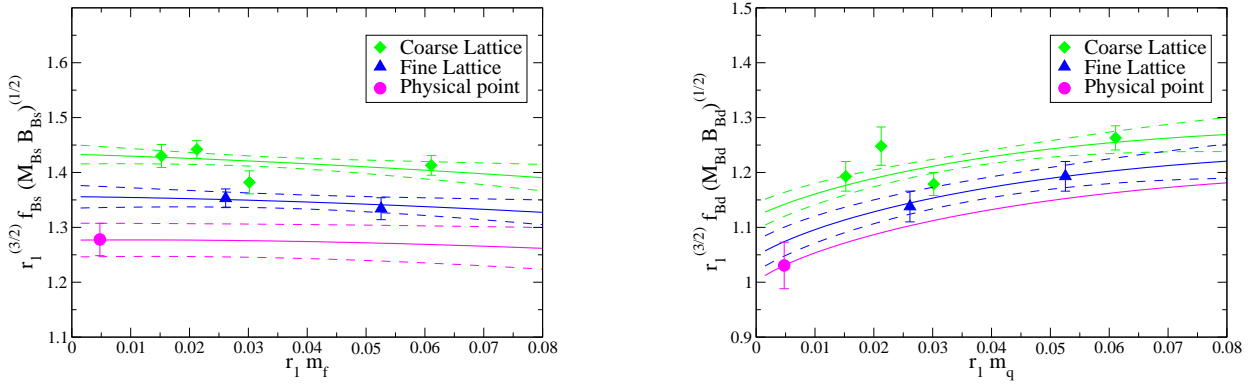

\centerline{
\ewxy{fbsr1.eps}{90mm}
\ewxy{fbdr1.eps}{90mm}
}
\caption{Chiral and continuum extrapolation of 
$\;r_1^{3/2} f_{B_s}\sqrt{M_{B_s}\hat{B}_{B_s}}$, 
$\;\;r_1^{3/2} f_{B_d}\sqrt{M_{B_d}\hat{B}_{B_d}}$. The red curve 
shows the continuum extrapolation and the ``physical point is at 
$m_{light}/m_s = 1/27.4$.}
\end{figure}

\begin{figure}
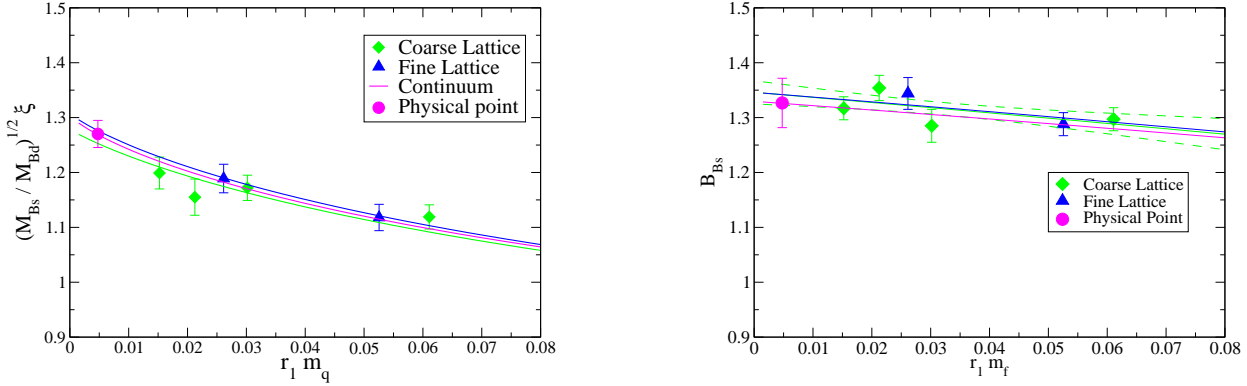

\centerline{
\ewxy{ximrat.eps}{90mm}
\ewxy{bagbs.eps}{90mm}
}
\caption{Same as Fig.1 for 
$\;\;\xi \sqrt{M_{B_s}/M_{B_d}}$ and for $\;\hat{B}_{B_s}$ }
\end{figure}

\section{Results}
\noindent
Using central values coming from the physical (red) points in the
 figures and the
errors summarized in Table I,  we can now present our main results \cite{bmix}.
\be
\label{xires}
\xi \equiv \frac{f_{B_s} \sqrt{B_{B_s}}}{f_{B_d} \sqrt{B_{B_d}}}
= 1.258(25)(21),
\qquad \hat{B}_{B_s} = 1.33(5)(3)
\ee
and using $r_1 = 0.321(5)fm$ \cite{upsilon},
\be
\label{fbsres}
f_{B_s} \sqrt{\hat{B}_{B_s}} = 266(6)(17) \;\left (\frac{0.321}{r_1[fm]} \right)^{3
/2}
{\rm MeV},
\qquad 
f_{B_d} \sqrt{\hat{B}_{B_d}} = 216(9)(12) \;\left (\frac{0.321}{r_1[fm]} \right)^{3
/2}
{\rm MeV },
\ee
where the first error comes from statistics + chiral extrapolation and
the second is the sum of all other systematic errors added in quadrature.
 The result for $f_{B_s} \sqrt{\hat{B}_{B_s}}$ in
eq.(\ref{fbsres}) is consistent with but more accurate than our
previously published  value of $281(21)$MeV \cite{bsmixing}.

\noindent
  Combining our lattice result for $\xi$ with the experimentally
measured mass differences $\Delta M_d = 0.507 \pm 0.005 \; ps^{-1}$ \cite{pdg} and
$\Delta M_s = 17.77 \pm 0.10 \pm 0.07 \; ps^{-1}$ \cite{cdf} leads to,
\be
\frac{|V_{td}|}{|V_{ts}|} = 0.214(1)(5)
\ee
where the first error is experimental and the second from the
lattice calculation presented here. Our Bag parameter result $\hat{B}_{B_s}$ 
can be combined with experimental $\Delta M_s$ and $\tau(B_s)$ to 
form the Standard Model prediction for the branching fraction 
$Br(B_s \rightarrow \mu^+ \mu^-)$ \cite{buras2}.  We find
\be
Br(B_s \rightarrow \mu^+ \mu^-)
 = 3.19(19) \times 10^{-9},
\ee
improving on the previous accuracy available.

\noindent
Our fits to a matrix of $B_q$ two-point correlators has also 
allowed us to update previous results on $B_d$ and $B_s$ decay 
constants. 
The final numbers  including all errors
 added in quadrature become,
\be
\label{fbrat}
\frac{f_{B_s}}{f_{B_d}} = 1.226(26),
\ee
\be
\label{fbd}
f_{B_d} =
 190(13) \;\left (\frac{0.321}{r_1[fm]} \right)^{3/2} {\rm MeV},
\qquad f_{B_s} =
 231(15) \;\left (\frac{0.321}{r_1[fm]} \right)^{3/2} {\rm MeV}.
\ee
These results for $f_{B_q}$ are consistent with but about one $\sigma$
lower than the values $f_{B_d} = 216(22)$MeV
and $f_{B_s} = 260(29)$MeV
given in \cite{fbprl, fbsprl}.
 The main difference between the analysis carried out
here and in \cite{fbprl} is that in the latter case chiral extrapolations were
done based only on coarse lattice data and furthermore no attempt
was made to extrapolate explicitly to the continuum limit.

\section{ $B_s$ Meson Decay Constant Using Other Actions}
The HPQCD collaboration has initiated a project of studying 
$B$ physics with NRQCD $b$-quarks and HISQ light quarks 
 \cite{egregory}.
 The hope is that by going from the AsqTad to the 
more highly improved HISQ 
action discretization errors will be further reduced, in particular 
those coming from taste breaking effects.  We are also simulating 
the $B_c$ system with NRQCD $b$- and HISQ charm-quarks \cite{bcstar}.
 Here we present first preliminary 
results for the $B_s$ meson decay constant $f_{B_s}$ in Fig.3.  
One sees that the slope versus lattice spacing is significantly 
reduced as one moves to the HISQ action.  Furthermore both actions 
lead to the same continuum limit within errors.  We are pursuing 
other approaches to $B$ physics as well, in particular one based 
on relativistic HISQ $b$-quarks. Initial calculations 
using relativistic $b$-quarks show results 
for $f_{B_s}$ in good agreement with the NRQCD-AsqTad and NRQCD-HISQ 
values presented here  \cite{efollana}.

\begin{figure}[b]
\begin{center}
\epsfysize=3.in
\centerline{\epsfbox{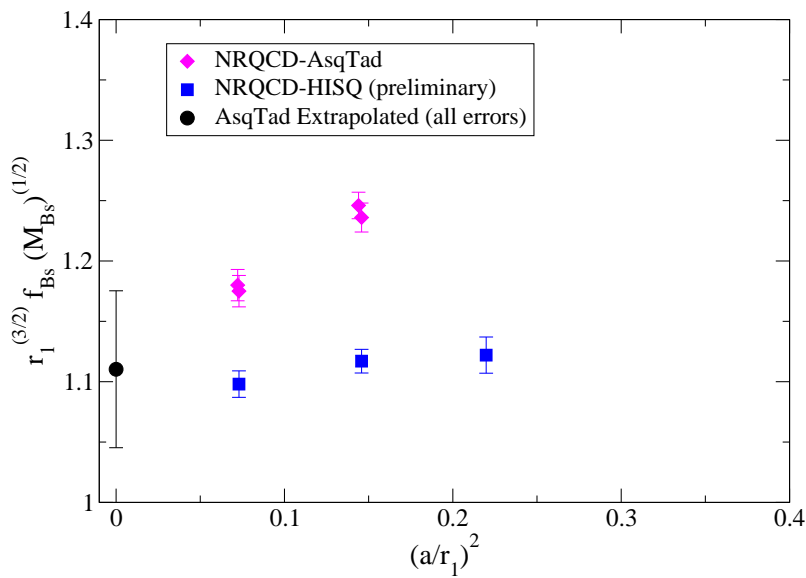}}
\end{center}
\caption{
Comparison between NRQCD-AsqTad and NRQCD-HISQ calculations 
of the $B_s$ meson decay constant.
  }
\end{figure}

\noindent
{\bf Acknowledgments}\\
The numerical simulations were carried out on facilities of the USQCD 
Collaboration funded by the DOE, at the Ohio Supercomputer Center and 
at NERSC.

\end{document}